\documentclass[journal]{IEEEtran}

\ifCLASSINFOpdf
\else
   \usepackage[dvips]{graphicx}
\fi
\usepackage{url}
\usepackage{color}
\usepackage{amsfonts}
\providecommand{\gz}[1]{\textcolor{black}{{#1}}}

\providecommand{\jd}[1]{\textcolor{black}{{#1}}}

\providecommand{\as}[1]{\textcolor{black}{{#1}}}

\usepackage{multirow}
\usepackage{amsmath}
\usepackage[table]{xcolor}
\usepackage{booktabs}
\usepackage{graphicx}

\begin{document}

\title{Music Source Separation with Generative Flow}

\author{Ge Zhu, Jordan Darefsky, Fei Jiang, Anton Selitskiy and Zhiyao Duan \IEEEmembership{Member, IEEE}
\thanks{This work was supported in part by National Science Foundation grants 1741472 and 1846184 and a New York State Center of Excellence in Data Science award.\\
Ge Zhu, Jordan Darefsky, Anton Selitskiy and Zhiyao Duan are with University of Rochester, Rochester, NY 14627, USA (e-mails:ge.zhu@rochester.edu, jdarefsk@u.rochester.edu,aselitsk@ur.rochester.edu,zhiyao.duan@rochester.edu). Fei Jiang was with University of Rochester. He is now with Tencent Technology (Shanghai) Co., Ltd., China (e-mail: flyjiang92@gmail.com).}}

\markboth{Journal of \LaTeX\ Class Files, Vol. 14, No. 8, August 2015}
{Shell \MakeLowercase{\textit{et al.}}: Bare Demo of IEEEtran.cls for IEEE Journals}
\maketitle

\begin{abstract}

Fully-supervised models for source separation are trained on parallel mixture-source data and are currently state-of-the-art. However, such parallel data is often difficult to obtain, and it is cumbersome to adapt trained models to mixtures with new sources. Source-only supervised models, in contrast, only require individual source data for training. In this paper, we first leverage flow-based generators to train individual music source priors and then use these models, along with likelihood-based objectives, to separate music mixtures. We show that in singing voice separation and music separation tasks, our proposed method is competitive with a fully-supervised approach. We also demonstrate that we can flexibly add new types of sources, whereas fully-supervised approaches would require retraining of the entire model.
\end{abstract}


\IEEEpeerreviewmaketitle

\section{Introduction}

Music source separation involves separating a music mixture into multiple source signals. It plays an important role in many downstream tasks~\cite{Manilow20ossbook} in music signal processing including melody extraction, lyric recognition and music search. 
Consequently, many algorithms have been proposed for various problem settings of music source separation in the past decades.


\jd{We categorize existing approaches as either supervised or unsupervised based on the availability of separated clean training data. An approach is supervised when any clean sources are available during training; an approach is unsupervised when no such data is available. Further, we define supervised approaches including the following two settings during training:}
\jd{
\begin{enumerate}
    \item \textbf{Fully-supervised} approaches: Both mixtures and their corresponding individual sources are available.
    \item \textbf{Source-only supervised} approaches: Only clean individual sources are available. This approach is sometimes referred to as unsupervised \cite{Narayanaswamy20unsupervised}, though we feel that learning to model source data is a form of supervision.
\end{enumerate}}
In recent literature, fully-supervised approaches achieve the state-of-the-art performance in source separation tasks. These approaches often involve training a model (e.g., a deep neural network) on parallel mixture-source data to map mixtures to their underlying sources (or their corresponding masks) \cite{mohammadiha2013supervised}. Such data can be naturally recorded or crafted to follow the real distribution of mixtures and their underlying sources. For example, the MUSDB18 dataset \cite{musdb18} consists of songs along with their corresponding individual sources. Such datasets, however, are difficult to construct or obtain. An alternative approach involves synthesizing training mixtures by randomly mixing clean training sources. This fully-supervised approach is referred to specifically as \textit{synthetic full-supervision}.

\jd{In the source-only supervised setting, a common approach involves first using the individual sources to learn models of the source domains and then using these models to perform separation of mixtures during inference. Notably, the non-negative matrix factorization~(NMF)~\cite{lee1999learning,virtanen2007monaural} based source separation is built upon the concept of a signal dictionary. Other approaches learn a probabilistic model~\cite{roweis2000one,benaroya2005audio,ozerov2007adaptation} for each source, and then separate the mixture with a signal reconstruction objective. More recently, the generative source separation framework~\cite{subakan2018generative} has gained much attention with the emergence of expressive deep generative models and various optimization techniques; for example, implicit generative models such as generative adversarial networks (GANs) have been used to train source priors~\cite{stoller2018adversarial,subakan2018generative,kong2019Single} which can then be used to separate sources with gradient-based methods \cite{subakan2018generative, Narayanaswamy20unsupervised}. Jayaram and Thickstun~\cite{jayaram2020source} train an explicit prior and sample with Langevin dynamics to perform source separation in the image domain; however, such sampling methods can be slow \gz{even with parallel sampling~\cite{jayaram2021parallel}}.}

\jd{In this paper, we focus on a source-only supervised, generative approach to music source separation. More specifically, we 1) train flow-based generators to model the spectrograms of various instruments; and 2) apply gradient-based optimization to separate sources at inference.}
Compared to fully-supervised methods, our approach only needs access to clean individual sources at train time; practically, it is easier to obtain individual source data than paired mixture-source data. \gz{Although synthetic full-supervision approach is shown to outperform traditional data augmentation~\cite{uhlich2017improving,defossez2019music} techniques, it requires a large amount of combinations of the sources~\cite{song2021catnet, kong2021decoupling}.} 
Compared to existing source-only supervised, generative methods, we find that using flow-based models provides two advantages in particular. First, flow-based models are invertible and thus have zero representation error; this is not the case for GAN-based generative priors~\cite{asim2020invertible}. This representation capability is beneficial for optimizing a reconstruction objective during separation. Second, we find empirically that the separation process converges quickly and that our approach is faster than current sampling-based methods.

On singing voice separation and music source separation tasks, we show that our proposed method outperforms current source-only separation approaches and achieves competitive performance with one of the fully-supervised methods. Furthermore, we demonstrate that we are able to flexibly add a new source. \gz{In contrast, in fully-supervised systems, to separate new sources, it is required to either alter the entire network architectures or prepare paired target source tracks and accompaniment tracks following one-versus-all training paradigm~\cite{stoter2019openunmix}.} We make the code\footnote{Open source code: \url{github.com/gzhu06/GenerativeSourceSeparation}} publicly available.

\section{Background}
\subsection{Generative source separation}
\label{sec:gss}
\jd{Source separation involves separating a mixture $\textbf{x}$ into $n$ individual sources $\textbf{s}_i$. Under an instantaneous mixing setting~\cite{ozerov2009multichannel}, we have: 
\begin{equation} 
\textbf{x} = \sum_{i=1}^n \alpha_i \textbf{s}_i, 
\end{equation}
\jd{where $\alpha_i$ is the mixing coefficient.} For simplicity, we assume $\alpha_i$=1. 
In probabilistic modeling framework~\cite{ozerov2007adaptation,subakan2018generative,jayaram2020source}, we can assume that different sources have different statistical behavior. Specifically,}
\begin{align}
    &\textbf{s}_i \sim p_G(\textbf{s}_i), \\
    &\textbf{x}|\textbf{s}_1,\ldots,\textbf{s}_n \sim p_{mix}\Bigl(\textbf{x}|\sum_{i=1}^n\textbf{s}_i\Bigr),
\label{eq:formulation}
\end{align}
\jd{where $p_G$ models sources $\textbf{s}_i$, and $p_{mix}$ models the mixture conditioned on the sum of sources. $p_{mix}$ is often fixed and chosen explicitly to model the noise between the sum of clean sources and the final mixture. }
\jd{To perform source separation, one could first train models $p_G$ to model the source distributions. Then, to perform separation, sources $\textbf{s}_i$ could be found to maximize the above maximum a posteriori (MAP) objective, using a preferred method of optimization. This method would be considered source-only supervised, as only individual sources are seen during training.}
\subsection{Flow models}

\as{Normalizing flow is a generative model that transforms a random variable $\mathbf{z}$ with a simple distribution $p_{\mathbf{z}}(\mathbf{z})$ (Gaussian in our case) into a target random variable $\mathbf{y}\sim p_{\mathbf{y}}(\textbf{y})$ through an invertible function $f_\theta$.} Using the change of variables formula, the log probability of $\mathbf{y}$ can be written as:
\begin{equation}
    \log p_{\mathbf{y}}(\mathbf{y}) = \log p_{\mathbf{z}}(\mathbf{z}) - \log\left|\det \frac{\partial f_\theta(\mathbf{z})}{\partial \mathbf{z}}\right|.
\label{eq:flowobjective}
\end{equation}
Often, $f_\theta$ is a composition of neural invertible flow layers, for which the Jacobians are efficient to compute. This allows for efficient computation of the total log-determinant term in Eq.~\eqref{eq:flowobjective}. Since $\textbf{z}$ is Gaussian, the $\log p_{\textbf{z}}(\textbf{z})$ term can directly be computed; thus, $f_\theta$ can be trained to maximize the log probability of data, $\log p_{\textbf{y}}(\textbf{y})$. In the case of source separation, we can use flow models as prior distributions $p_G$.

\section{Proposed Method}

\begin{figure*}[t]
\centering
\includegraphics[width=4.2in]{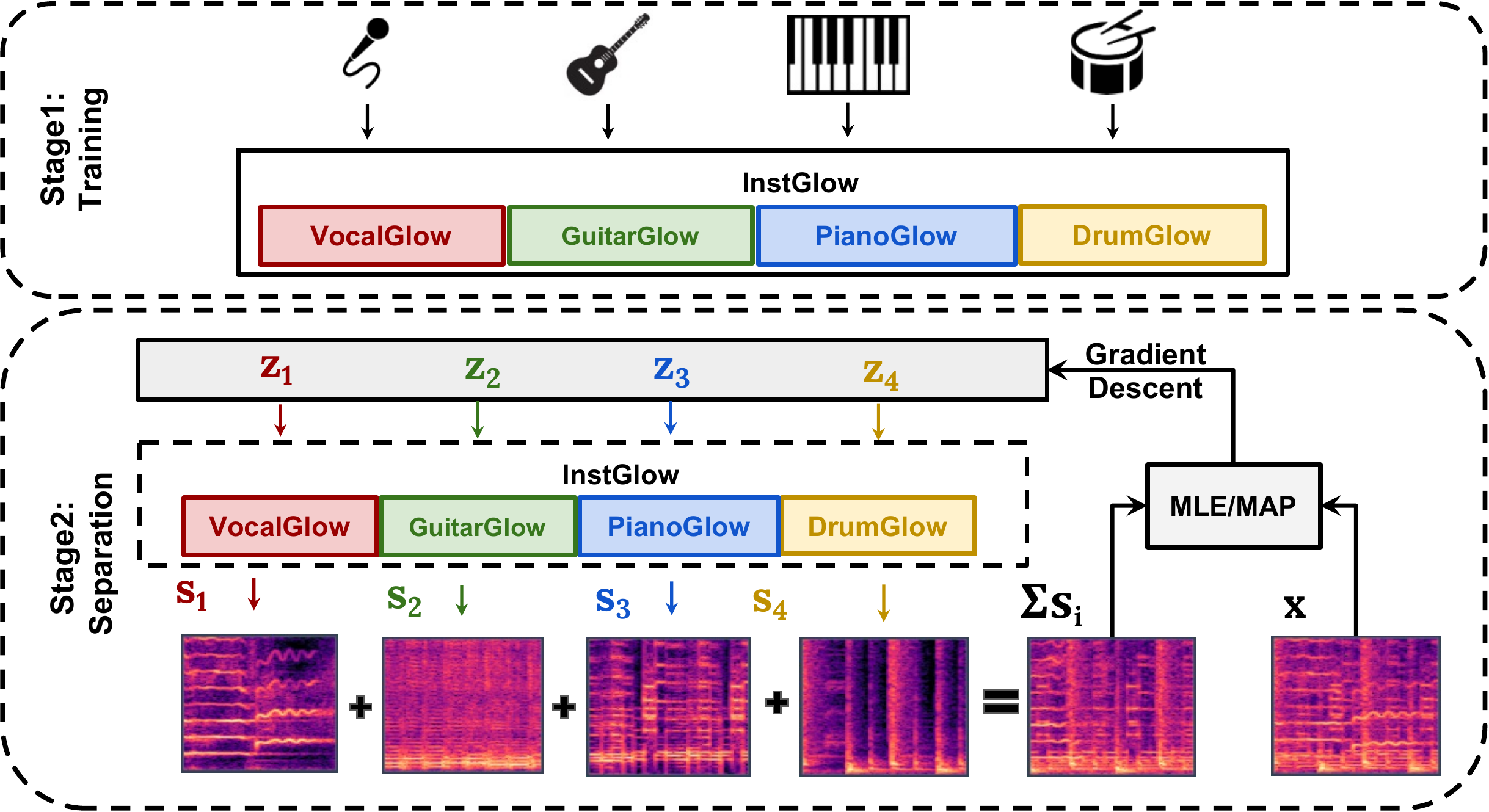}
\caption{Diagram for proposed flow-based generative source separation. Stage 1: training source prior models with instrument-specific unconditional models (\textit{InstGlow}), one for each source. Stage 2: separating sources by searching the optimal latent code $\{z_i\}$ to optimize an MLE or an MAP objective.}
\label{fig:dia}
\end{figure*}

\subsection{Glow priors}


Having witnessed the success in solving inverse problems with flow-based 2D image priors~\cite{asim2020invertible}, we use Glow~\cite{kingma2018glow} as our generative model backbone to learn source priors from 2D audio magnitude spectrograms. One motivation for modeling music priors in the spectral domain is that the magnitude spectrograms of singing voice and background music have different structures, which may facilitate separation;  singing voice spectrograms tend to be sparse while background music spectrograms tend to be low rank and change more slowly~\cite{huang2012singing}. Note that the magnitude spectrogram of the mixture is not the exact sum of those of the sources due to phase differences; however, the sum is a good approximation as shown in NMF-based methods~\cite{virtanen2008bayesian}.

Fig.~\ref{fig:dia} illustrates the training and inference (i.e., separation) process of our proposed flow-based model. For the task of music source separation, we train a set of independent Glow priors (named \textit{InstGlow}), one for each source. 
 


\gz{We adapt the Glow~\cite{kingma2018glow} as the flow-based generator backbone} and use $\textbf{z}_i\sim \mathcal{N}(0,\textbf{I})$ as the latent prior. Our glow generator consists of a squeeze layer, 12 flow blocks, and an unsqueeze layer. The squeeze and unsqueeze operation follows the design in~\cite{kim2020glow}. In each step of flow, we use an activation normalization layer, an invertible 1x1 convolution layer~\cite{kim2020glow}, and an affine coupling layer in~\cite{prenger2019waveglow} without local conditioning.

\begin{figure}[t]
\centering
\includegraphics[width=0.84\columnwidth]{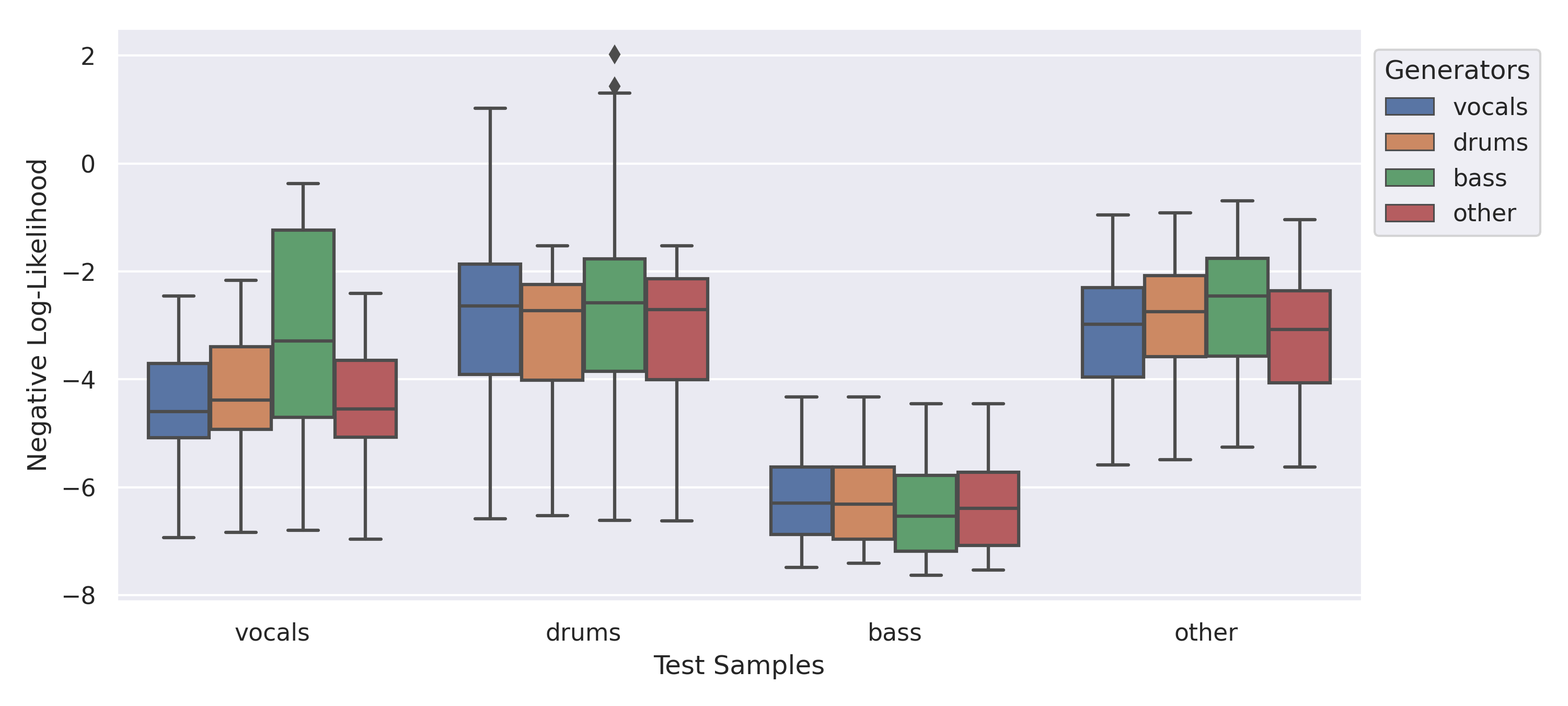}
\caption{\gz{Boxplots of NLLs of pre-trained `vocals', `drums', `bass' and `other' Glow generators (different colors) on a total of 100 one-minute test audio pieces from the four source categories (different columns). Each data point is the NLL of one generator on one test audio piece.}}
\label{fig:llfig}
\end{figure}
\subsection{Inference}
In the separation stage, \gz{we assume that we have knowledge of sources presented in the mixture and apply \textit{all} of the predefined source priors to separate corresponding components.} As mentioned in section ~\ref{sec:gss}, we use MAP~\cite{benaroya2005audio} as the separation objective and apply \gz{an iterative optimization} to separate predefined sources:


\begin{align}
   \hat{\textbf{s}}_1,\ldots,\hat{\textbf{s}}_n&= \underset{\textbf{s}_1,\ldots,\textbf{s}_n}{\operatorname{argmax}}~\log p(\textbf{s}_1,\ldots,\textbf{s}_n|\textbf{x}) \\ &=\underset{\textbf{s}_1,\ldots,\textbf{s}_n}{\operatorname{argmax}} ~ \log p(\textbf{x}|\textbf{s}_1,\ldots,\textbf{s}_n)+  \sum_{i=1}^n \log p(\textbf{s}_i),
\label{eq:map}
\end{align}
where $\textbf{x}$ is the observed mixture. In Eq. (\ref{eq:map}) we assume \gz{statistical} independence among all source tracks. In the above MAP formulation, we can also optimize over latent variable $\textbf{z}_i$ rather than $\textbf{s}_i$, as there is a bijection between them from the Glow model, $\textbf{s}_i =f^i_\theta(\textbf{z}_i)$.

To model $p(\textbf{x}|\textbf{s}_1,\ldots,\textbf{s}_n)$ \gz{in the instantaneous mixing}, we assume an independent additive \gz{residual} noise $\textbf{n}$ over the sum of the sources~\cite{benaroya2005audio}, i.e., 
$p(\textbf{x}|\textbf{s}_1,\ldots,\textbf{s}_n)=p(\textbf{x}-\sum_i^n \textbf{s}_i)= p(\textbf{n})$.
We assume that \gz{the spectrogram magnitude of the residual noise follows 
a Poisson distribution, then the log-likelihood of the mixture (first term in Eq.~(\ref{eq:map})) becomes \as{equivalent to} the negative generalized KL-divergence.}
Because we are using flow-based models, the second term in Eq.~(\ref{eq:map}) (the exact log-likelihood of the source priors) can be computed directly. 
We optimize the objective Eq.~\eqref{eq:map} \gz{by searching the latent space within the support of pre-trained generators~\cite{asim2020invertible,whang2021solving}}. Since both terms in Eq.~(\ref{eq:map}) are differentiable with respect to $\textbf{z}_i$, we perform this optimization with gradient descent. \gz{During optimization, we initialize $\textbf{z}_i=0$ to bias the latent codes towards zero in order to align with the target $\textbf{z}_i$ after the source specific priors are trained. This bias could be viewed as a simpler prior, with the benefit of being more robust to high out-of-distribution likelihoods.} After finding \gz{the optimal} latent codes $\textbf{z}_i$, we can compute the spectrograms of sources using the Glow models with $\textbf{s}_i =f_\theta^i(\textbf{z}_i)$. Eventually, we synthesize the source waveforms using inverse-STFT with the \gz{recovered source spectrogram} and the mixture phase.
\subsection{Prior reweighting}
Previous works on speech enhancement~\cite{shi2021can}, audio source separation~\cite{frank2020problems}, and image inpainting tasks~\cite{asim2020invertible, whang2021solving} have found that flow-based models tend to assign high probability density to some out-of-distribution data while assigning low density to some in-distribution data. We find similar phenomena in our experiments. We computed the negative log-likelihoods (NLLs) of the four pretrained Glow priors on the 100 one-minute source tracks from the test partition in MUSDB18 shown in Fig.~\ref{fig:llfig}. We observe that the estimated NLLs are highly correlated and overlapping with each other for the same samples, suggesting that the pre-trained instrument generators are not discriminative enough in differentiating unseen instruments at inference.

To address this concern, we empirically re-weigh the prior term in Eq.~\eqref{eq:map} with coefficient $\gamma\in[0,1]$, initially proposed in~\cite{asim2020invertible}. We can discard the prior term in Eq.~\eqref{eq:map} by choosing $\gamma=0$ and arrive at a maximum likelihood estimation (MLE) objective. \gz{Notice that, we keep the zero initialization of $\textbf{z}_i$ in the MLE approach to avoid trivial solutions for Eq.~(\ref{eq:map}) without the prior term constraints.} Also note that in this MLE objective we effectively treat our Glow model as an implicit generator~\cite{mohamed2016learning}, though in our case sources are deterministically related to the latents.


\section{Experiments}
\begin{table*}[t]
\centering
\caption{Comparison of source separation systems with median SDR (dB) across tracks on three settings of two test sets. Grey cells indicate that the system is unable to separate that source type unless it is \gz{retrained from scratch using one-versus-all paradigm} and on the same kind of sources as the test set.}
\begin{tabular}{cccccccccccc}
\toprule
\multirow{2}{*}{\textbf{Method}} & \multicolumn{1}{c}{\multirow{2}{*}{\parbox{1.5cm}{\centering \textbf{Neural Networks}}}} &
\multirow{2}{*}{\textbf{Supervision}}& \multicolumn{5}{c}{MUSDB18-22.05kHz~\cite{musdb18}} & \multicolumn{4}{c}{Slakh2100-submix~\cite{manilow2019cutting}}\\
\cmidrule(lr){4-8}\cmidrule(lr){9-12}
 &&& {Vocals} & {Bass} & {Drums} & {Other} & {Acc.}& {Bass} &{Drums}&{Guitar} & {Piano} \\
\midrule
\textsc{InstGlow-MLE (Ours)} & Glow&Source-only& \textbf{3.92} & \textbf{2.58} &\textbf{3.85}& \textbf{2.37}&\textbf{9.82} & \textbf{1.54}  & \textbf{6.14} & \textbf{1.85} & \textbf{0.80}  \\
\textsc{InstGlow-MAP (Ours)} & Glow&Source-only& 3.66 & 2.51 &3.70& 1.99&9.52 & 1.39  &5.95 & 1.51 & 0.51 \\
\midrule
\textsc{LQ-VAE~\cite{mancusi2021unsupervised}}  & VQ-VAE &Source-only& 0.16  & {\cellcolor{black!5} --}  & {\cellcolor{black!5} --}  & {\cellcolor{black!5} --} & 4.47  & {\cellcolor{black!5} --}  & {\cellcolor{black!5} --}  & {\cellcolor{black!5} --}  & {\cellcolor{black!5} --}   \\
\textsc{GAN-prior~\cite{Narayanaswamy20unsupervised}} & SpecGAN&Source-only & -0.44 & 0.48 &-0.40& 0.32&4.29 & 0.09   & 0.85 & -0.01 & -0.42\\
\midrule
Conv-TasNet~\cite{luo2019convtas} & TCN & Full & 7.00  & 4.19 &5.25& 3.94&12.84 & 4.97 & 9.95 &{\cellcolor{black!5} --}   & {\cellcolor{black!5} --}  \\
Demucs (v2)~\cite{defossez2021hybrid}  & U-Net &Full & \textbf{7.14} & \textbf{5.50} & \textbf{6.74} &\textbf{4.16}&\textbf{12.94}& \textbf{5.48}  &\textbf{10.21} &{\cellcolor{black!5} --} & {\cellcolor{black!5} --}   \\
Open Un-mix ~\cite{stoter2019openunmix}& BiLSTM &Full & 6.86 &4.88 &6.35& 3.86&12.75& 4.66& 8.64  & {\cellcolor{black!5} --} & {\cellcolor{black!5} --}    \\
Wave-U-Net~\cite{stoller18waveunet}  & Wave-U-Net&Full& 5.06 &2.63 &3.74& 1.95&7.02& 0.01 &3.91 & {\cellcolor{black!5} --} & {\cellcolor{black!5} --}    \\
\bottomrule
\end{tabular}

\label{tab:music_sep_results}
\end{table*}
\label{sec:exp}
\subsection{Dataset}

We train the \gz{source priors for \texttt{vocals}, \texttt{bass}, \texttt{drums} and \texttt{other}} using the train subset of the MUSDB18 \gz{and \texttt{guitar} and \texttt{piano} from the train subset of Slakh2100 (i.e. we do not use \texttt{bass}, \texttt{drums} source tracks from Slakh2100 to train the source priors)}. For preprocessing, we \gz{use the mono channel and} downsample the tracks into 22.05 kHz and split them into 5-second non-silent segments. We use spectrograms with 1024-point FFT size and 256-point hop size as input features.

For MUSDB18 evaluation, we test both multi-instrument separation and singing-accompaniment separation. To construct accompaniment tracks, we sum the separated non-vocal sources. For Slakh2100-submix evaluation, we select and remix the subtracks of the top four instrument categories (\texttt{piano}, \texttt{bass}, \texttt{guitar} and \texttt{drums}) from the original Slakh2100.
We split the test portion into one-minute segments to fit into memory. We measure the global signal-to-distortion ratio (SDR) defined in Music Demixing Challenge~\cite{mitsufuji21music} of each segment with museval toolbox~\cite{stoter20182018} to evaluate separation performance. Following~\cite{mitsufuji21music}, we remove silent segments in the test data, where SDR is undefined.

\subsection{Baselines and training}
We use Conv-TasNet, Demucs(v2)~\cite{defossez2019music}, open-unmix~\cite{stoter2019openunmix} and Wave-U-Net~\cite{stoller18waveunet} as fully-supervised baseline systems. We directly use the authors' pre-trained models trained on MUSDB18. Since MUSDB18 does not contain guitar and piano source tracks, these models cannot separate such sources in Slakh2100-submix without \gz{preparing paired source-mixture data and retraining from scratch.}

We also compare our model to source-only supervision systems, using GAN-prior~\cite{Narayanaswamy20unsupervised} and LQ-VAE~\cite{mancusi2021unsupervised} as baselines. 
During separation, we apply projected gradient descent (PGD) on the reconstruction objective to search for the latent codes. For LQ-VAE, we apply the authors' method as-is.
\gz{For source priors training, we use the Adam~\cite{kingma2014adam} optimizer at a learning rate of 1e-4 for 1000 epochs.} During separation, we use the Adam optimizer at a learning rate of 0.01 for 150 iterations. Each iteration takes 0.3 seconds during evaluation on NVIDIA 2080Ti GPU.
We also conduct an ablation study on MLE and MAP optimization objectives.

\subsection{Results}

We start by comparing different variants of our proposed method, shown in the first two rows in Tab.~\ref{tab:music_sep_results}. We observe that the InstGlow with the MLE objective achieves the best results in terms of SDR across all of the tasks. For the InstGlow models, the MLE objective that only uses KL-divergence achieves better performance than MAP estimation; this differs from results in image domain ~\cite{whang2021solving}, where MAP estimation shows better performance. One reason may be due to the independence assumption of instrument sources in MAP estimation in Eq.~(\ref{eq:map}). Another reason may result from the fact that deep generative models lack of discriminative abilities to distinguish data of other classes
~\cite{nalisnick2018deep}, especially for the high dimensional data~\cite{nalisnick2018deep}. 

\jd{When comparing to source-only supervision systems shown in the middle rows of Tab.~\ref{tab:music_sep_results}, our proposed InstGlow significantly outperforms the other systems. While the results for the GAN-prior~\cite{Narayanaswamy20unsupervised} are perhaps surprisingly poor, our findings are consistent with the authors of LQ-VAE, who report that the GAN-prior performs poorly on the drum-piano toy dataset. We suspect the LQ-VAE baseline performs poorly due to the relatively small dataset (MUSDB18) on which we trained it. Note that in a similar method to LQ-VAE, \cite{el2022transfer, manilow2022source} uses a pre-trained Jukebox VQ-VAE-model (on 1.2 million songs) \cite{dhariwal2020jukebox}, \gz{and El Amri~\textit{et al.} \cite{el2022transfer}} achieved comparable performance to fully-supervised methods. However, for a fair comparison with our method and the GAN-prior baseline, we only pre-train LQ-VAE on the MUSDB dataset. Also note that LQ-VAE is only able to separate two sources, due to its training paradigm.} 

We compare our best performing system, InstGlow-MLE, with the fully-supervised baselines. On the MUSDB18 test set, \gz{a statistical test shows that InstGlow-MLE significantly outperforms Wave-U-Net in other, and achieves comparable results in bass and drums, although there is still a large gap behind Wave-U-Net in vocals.} \gz{By listening to the separated vocal samples from InstGlow-MLE, we found that they contain more interference from other sources, which could be eased by adding regularization such as coherence loss~\cite{tian2020deep}.}

On the Slakh2100-submix test set, we can only compare with fully-supervised models on the bass and drums sources, as their models are trained on MUSDB18 which does not contain guitar and piano tracks. We observe that InstGlow-MLE achieves better results than Wave-U-Net \gz{without retraining on bass and drums sources from Slakh2100 dataset}, showing its generalization ability to new datasets. 
Finally, we additionally train guitar and piano priors using \gz{\textit{only}} the source data from the Slakh2100 training subset and apply them to separate the corresponding tracks in the Slakh2100-submix test subset. We find that the separation performance of guitar and piano is similar to that of the bass source but lower than that of the drum source; this relatively weak performance may be due to the significant pitch range overlap of guitar and piano~\cite{manilow2019cutting}. \gz{We find this training paradigm promising given the above observations, as InstGlow only requires instrument data and can be used to separate sources undefined in MUSDB18, which is not easily feasible in fully-supervised models.}


\section{Conclusions}

In this paper, we employed flow-based generators for music source separation in the source-only supervision setting. To the best of our knowledge, we are the first to report successful separation results in this setting on benchmark separation tasks, achieving significantly better results than other source-only supervised methods. Future work is to bridge the performance gap between our method and fully-supervised approaches \gz{by potentially scaling up with more instrument data in the wild} as well as to extend it to more general settings. 


\bibliographystyle{IEEEtran}
\bibliography{mybib}

\end{document}